# Ta$_4$CoSi: a Tantalum-rich superconductor with a honeycomb network structure

*Lingyong Zeng[a], Xunwu Hu[b], Shu Guo[c,d], Gaoting Lin[e], Jing Song[f], Kuan Li[a], Yiyi He[a], Yanhao Huang[a], Chao Zhang[a], Peifeng Yu[a], Jie Ma[e], Dao-Xin Yao[b,d], Huixia Luo[a]\**

[a]School of Materials Science and Engineering, State Key Laboratory of Optoelectronic Materials and Technologies, Key Lab of Polymer Composite & Functional Materials, Guangzhou Key Laboratory of Flexible Electronic Materials and Wearable Devices, Sun Yat-Sen University, No. 135, Xingang Xi Road, Guangzhou, 510275, P. R. China

[b]School of Physics, Center for Neutron Science and Technology, Guangdong Provincial Key Laboratory of Magnetoelectric Physics and Devices, State Key Laboratory of Optoelectronic Materials and Technologies, Sun Yat-Sen University, Guangzhou, 510275, P. R. China

[c]Shenzhen Institute for Quantum Science and Engineering, Southern University of Science and Technology, Shenzhen 518055, China

[d]International Quantum Academy, Shenzhen 518048, China

[e]Key Laboratory of Artificial Structures and Quantum Control, Shenyang National Laboratory for Materials Science, School of Physics and Astronomy, Shanghai Jiao Tong University, Shanghai 200240, China.

[f]Institute of Physics, Chinese Academy of Sciences, Beijing 100190, China.

*Corresponding author/authors complete details (Telephone; E-mail:) (+86)-2039386124; E-mail address: luohx7@mail.sysu.edu.cn;

**Abstract**

The crystal structure of solid-state materials with unique lattices is considered one of the main factors determining physical properties, including superconductivity. Materials with honeycomb lattices have inspired intense research interest for their novel properties. A previously unknown compound, $Ta_4CoSi$, are discussed herein and its crystal structure and superconducting properties. It crystallizes in a $CuAl_2$-type structure (space group *P4/mcm*, No. 124) with the lattice parameters $a = b = 6.17755(7)$ Å and $c = 4.98443(8)$ Å, featuring honeycomb networks of Ta-Ta in (110) plane. Superconductivity is observed below the critical temperature of ~ 2.45 K, while lower and upper critical magnetic fields are ~ 9.86 mT and 0.84 T, respectively. Experiments and theoretical calculations show that the honeycomb lattices significantly influence the superconducting properties. This material may thus provide a new platform for investigating the exotic superconductivity of honeycomb networks.

## I. Introduction

Ternary silicide intermetallic compounds $A_xB_y\text{Si}_z$ ($A$, $B$ = transition metal or rare-earth metal) have shown very diverse and interesting crystal structures and physical properties. Depending on the constituent element, properties like superconductivity, heavy-fermion, quantum spin liquids, large magnetocaloric effect, large magnetoresistance, etc. have been observed in this family [1-11]. The search for new superconductors in ternary silicide intermetallic systems has always been one of the research hotspots in the field of condensed matter. And many novel superconductors have been reported in this family, such as a wide variety of non-centrosymmetric superconductors ($AB\text{Si}_3$) [1,6-8], multi-gap superconductors ($\text{Lu}_2\text{Fe}_3\text{Si}_5$ and $\text{Sc}_5\text{Ir}_4\text{Si}_{10}$) [2,3], the first heavy fermion superconductor ($\text{CeCu}_2\text{Si}_2$) [4], and the strong correlation superconductor ($\text{Er}_2\text{Fe}_3\text{Si}_5$) with the coexistence of antiferromagnetism and superconductivity [5]. Besides, equiatomic ternary silicide intermetallic ($AB\text{Si}$) are noticeable, which are well-known superconductors with a different superconducting transition temperature ($T_c$) [12-16].

Furthermore, another family of layered materials with a Kagome structure also attracts attention. Recently, the ternary silicide intermetallic $\text{LaRu}_3\text{Si}_2$ is a nodeless moderate coupling Kagome superconductor, with the $T_c \sim 7.3$ K [17,18]. The structure of $\text{LaRu}_3\text{Si}_2$ contains distorted Kagome layers of Ru sandwiched between La layers and Si layers with a honeycomb network. Meanwhile, honeycomb lattices are also an attractive playground for superconductivity. Although the superconducting compounds in ternary silicide intermetallic systems are abundant, there is still a high potential to investigate and discover novel types of superconductors in these systems.

$\text{Nb}_4M\text{Si}$ (where $M$ = Fe, Co, or Ni) are the first representatives of ternary silicide intermetallic compounds with a structure of the $\text{CuAl}_2$-type. Two kinds of space groups of $P4/mcc$ and $I4/mcm$ were previously reported for these $\text{CuAl}_2$-type compounds [19]. And previous studies have demonstrated that the $P4/mcc$ space group is a superconducting phase. In addition, there are few reports on superconducting cobalt-based compounds. Nb$_4$CoSi is one of the cobalt-based superconductors ($T_c$ = 6.0 K) that includes Co as a principal constituent element [20]. And an interesting feature is that although binary compounds of the CuAl$_2$-type are unknown for niobium, tantalum forms the compounds Ta$_2$Si and Ta$_2$Co exhibiting structures of this type. It is possible that these compounds will present CuAl$_2$-type structure in the ternary Ta-Co-Si system, which has not yet been studied, and that restricted solid solutions of cobalt in Ta$_2$Si will be found in the ternary Ta-Co-Si systems.

In this paper, we have successfully synthesized a new intermetallic compound in the Ta-Co-Si ternary system. The Ta$_4$CoSi has a Ta-Ta honeycomb network in the (110) plane. It displays superconductivity with $T_c$ = 2.45 K. Electric resistivity, magnetic susceptibility, and specific heat measurements have been carried out to determine the Ta$_4$CoSi alloy's superconducting parameters. First-principles calculations indicate that the Ta 5$d$ state mainly dominates the electronic structure of the Ta$_4$CoSi intermetallic compound.

## II. Experimental Details

Elemental Ta (99 %, 22 mesh, Alfa Aesar), Co (99.5 %, 325 mesh, Alfa Aesar), and Si (99.9 %, 100 mesh, Alfa Aesar) for Ta$_4$CoSi were utilized as raw materials and synthesized by the arc-melting method. 250 mg of the elemental powders in a molar ratio of 3.2:1:1 was pressed into a piece, with Ta powder on the top and Si powder on the bottom, then arc-melted under 0.5 atmosphere of argon gas. The ingots were turned over and melted several times from both sides of the ingot to ensure homogeneity. The resulting air-stable material was annealed at 1050 ºC for 48 h under a high vacuum.

The crystal structure of the sample was determined by the powder X-ray diffraction (PXRD) using a conventional X-ray spectrometer equipped with Cu K$a$ radiation. The intensity data were obtained over the 2$\theta$ range of 10 - 100º with a step width of 0.01º. The lattice parameters and the occupancy of the atoms were refined using the Rietveld method. The Fullprof suite package was used for Rietveld fitting of the PXRD data [21]. The elemental ratio was determined by scanning electron microscope combined with energy-dispersive X-ray spectroscopy (SEM-EDXS). Electrical transport, magnetization, and specific heat measurements were measured in a Quantum Design PPMS-14T. The resistivity was performed with the standard four-probe configuration by attaching Pt electrodes with silver paste on the sample.

We performed first-principles calculations implemented in Vienna Abinit Simulation Package (VASP) [22, 23]. The generalized gradient approximation (GGA) in the Perdew-Burke-Ernzerhof (PBE) form was used in treating the electron exchange-correlation [24]. The pseudopotentials for the projector augmented wave (PAW) method (where valence electrons configurations of each atom are $3d^84s^1$ for Co, $3s^23p^2$ for Si, and $5p^66S^25d^3$ for Ta) with basis set cutoff energy of 500 eV were used to describe the ion-electron interaction [25]. The lattice constants of polycrystalline refined from X-ray diffraction were fixed for comparison with the experiment. In contrast, atomic positions were fully relaxed until the Hellmann-Feynman force on each atom was less than 0.01 eV/Å in the ionic relaxation loop. The k-point sampling of the first Brillouin zone (BZ) is 5×5×7 Γ-centered K-mesh to ensure convergence.

**III. Results and Discussion**

PXRD characterization of the ground-polycrystalline Ta$_4$CoSi is shown in **Fig. 1a**. Ta$_4$CoSi crystallizes in a CuAl$_2$-type structure, the same as other reported 411-phases [20,26] and the recent high entropy alloys superconductors [27]. Consistent with Nb$_4$NiSi [19,20], there may also exist two space groups *P4/mcc* and *I4/mcm* in Ta$_4$CoSi compound. And the diffraction peaks of these two space groups basically coincide. The difference between these two space groups is whether Si and Co atoms are ordered or disordered. When refining the PXRD data only with *P4/mcc* space group, the refinement parameters $R_{wp}$, $R_p$, and $\chi^2$ are 4.51%, 3.23%, and 7.26, respectively. However, we could obtain smaller refinement parameters by applying two space group model (see in **Fig. 1a**). And the refinement parameters $R_{wp}$, $R_p$, and $\chi^2$ are 2.65 %,

3.23 %, and 5.15, respectively. The fitting results show that the main phase is about 76 % of the *P4/mcc* phase and about 15 % of the *I4/mcm* phase. Some extra peaks originate from the excess Ta in the raw material. The lattice parameters of *P4/mcc* space group in $Ta_4CoSi$ from Rietveld fitted PXRD profile are $a = b = 6.17755(7)$ Å, $c = 4.98443(8)$ Å. **Table 1** lists the refined structural parameters of *P4/mcc* phase. The refined crystal structure with *P4/mcc* space group of $Ta_4CoSi$ is shown in **Fig. 1b**. As shown in **Fig. 1b**, the nearest Ta atoms form a one-dimensional zigzag along the c-axis. The nearest Co and Si atoms form linear chains along the c-axis. The Ta atoms in $Ta_4CoSi$ is stretched to the direction of the (110) plane of the crystal structure, forming a hexagonal Ta-chain network, similar to graphite, with two different Ta-Ta band length (2.80 Å and 2.94 Å). The atom having a honeycomb structure is also observed in $MgB_2$ [28], non-centrosymmetric $TaRh_2B_2$ [29], and Kagome lattice superconductors [30]. The back-scattered electron (BSE) and EDXS were characterized to check the homogeneity and the chemical composition of $Ta_4CoSi$. As displayed in **Fig. 2**, all constituent elements distribute homogenously, and the real chemical formula is $Ta_{4.05}Co_{0.89}Si_{1.06}$.

The temperature-dependent electrical resistivity from 300 to 1.8 K of the polycrystalline $Ta_4CoSi$ sample is presented in **Fig. 3**. The sample displays metallic behavior. The resistivity changes slightly above $T_c$, and the residual resistivity ratio RRR = $\rho(300 K)/\rho(5 K) \approx 1.25$, is a low ratio, comparable to that observed for highly disordered intermetallic or nonstoichiometric compounds. The inset in **Fig. 3** displays the low-temperature behavior in the vicinity of $T_c$. The onset temperature of $T_c$ appears at about 2.52 K, whereas the zero resistivity is reached at about 2.41 K, characteristic of superconductivity. The sharp resistivity transition covers only a 0.11 K temperature range. The $T_c \approx 2.45$ K is determined from the midpoint of the resistive transition. The zero-field cooled (ZFC) magnetization of the $Ta_4CoSi$ is displayed in **Fig. 4a**. Measurements were carried out between 1.8 K and 10 K under a 20 Oe field. The field-dependent volume magnetization was performed from 0 to 200 Oe for $Ta_4CoSi$ at 1.8 K and fitted by the formula $M_{fit} = a + bH$, where b represents the slope of the fit line.

The slope of the linear fit is -0.11082. The procedure, $-b = 1/(4\pi(1-N))$, was then used to estimate the demagnetization factor, $N$, which depends on the shape of the sample and how the sample is oriented concerning the magnetic fields. And the resulting $N$ is about 0.28 for $Ta_4CoSi$ for the measured sample, close to the theoretical value (0.33) for a rectangular cuboid sample [31]. The actual value of $N$ (0.28) has been used to correct the magnetic susceptibility data, $4\pi\chi_v(1-N) = 0.78$. The PXRD refinement indicates that the sample has a superconducting phase (*P4/mcc*) of about 74 %, which is close to the magnetic susceptibility data (78 %). The strong diamagnetic response at 1.8 K approaches the ideal value of $4\pi\chi_v(1-N) = -1$ (-1 represents the ideal value of fully superconducting materials) for $Ta_4CoSi$ superconductor, implying bulk superconductivity. The critical temperature of $T_c = 2.44$ K was determined as the value at the point where the linearly approximated slope crosses the normal state magnetization, seen in the arrow in **Fig. 4a**.

The new Ta-rich superconductor $Ta_4CoSi$ was further performed with field dependence of magnetization measurement at various temperatures below $T_c$, as shown in **Fig. 4b**. With increasing magnetic fields, magnetization rapidly decreases. The low-field linear fitted the magnetization data ($M_{fit}$), as previously described, were used to construct the M-$M_{fit}$ plot in **Fig. 4c**. The field where the magnetization starts to deviate from the linear response is the uncorrected lower critical field, $\mu_0 H_{c1}^*$ at that temperature. All values with the corresponding temperatures are displayed in **Fig. 4d** and fitted to the below formula: $\mu_0 H_{c1}^*(T) = \mu_0 H_{c1}^*(0)(1-(T/T_c)^2)$, which is illustrated by the solid purple line. The $\mu_0 H_{c1}^*(0)$ was calculated to be 7.10 mT. Furthermore, when considering the $N$ value 0.28, the lower critical field is calculated by the equation: $\mu_0 H_{c1}(0) = \mu_0 H_{c1}^*(0)/(1-N)$. Therefore, the corrected lower critical field $\mu_0 H_{c1}(0)$ is 9.86 mT, which is somewhat smaller than that of the $Nb_4NiSi$ sample (10.9 mT) [20].

To further understand the superconducting state, the magneto-resistivity and the magnetic field-dependent magnetization under different temperatures were measured. The resistivity under different magnetic fields is presented in **Fig. 5a**. $T_c$ at various magnetic fields is determined at the 50 % drop from the normal state resistivity value. **Fig. 5b** presents the $H_{c2}$-T phase diagram of the superconducting states for Ta$_4$CoSi alloys. The resulting slope ($d\mu_0H_{c2}/dT$) is -0.4952 T/K for Ta$_4$CoSi. The upper critical field $\mu_0H_{c2}(0)$ can then be obtained via the Werthamer-Helfand-Hohenberg (WHH) formula: $\mu_0H_{c2}(0) = -0.693T_c(\frac{d\mu_0H_{c2}}{dT})|_{T=T_c}$. Using $T_c$ = 2.45 K, the dirty limit $\mu_0H_{c2}(0)^{WHH}$ value was calculated to be 0.84 T. And the Pauli limit transition field $\mu_0H^{Pauli} = 1.85*T_c$ = 4.53 T.

After observing both the zero resistivity and the Meissner effect in the magnetization, to confirm the observed superconductivity is intrinsic in Ta$_4$CoSi, specific heat measurements were carried out from 0.6 to 10 K, as shown inset in **Fig. 6**. An apparent anomaly at $T_c$ = 2.34 K in the zero-field heat capacity corresponds to the emergency of superconductivity. It is well consistent with the $T_c$ determined from resistivity and magnetization measurements. The data were fitting with the equation $C_p/T = \gamma + \beta T^2$, yielding the electronic specific heat coefficient $\gamma$ = 15.1 mJ/mol/K$^2$ and the phonon specific heat coefficient $\beta$ = 0.28 mJ/mol/K$^4$. After subtracting the phonon contribution, the electronic specific heat ($C_{el}$) is isolated and plotted as $C_{el}/\gamma T$ against $T/T_c$ in Fig. 6. The $C_{el}/\gamma T$ jumps are analyzed by a modified BCS model, the so-called $\alpha$ model [31]. Using the equal area construction method, the normalized heat capacity jump ($\Delta C_{el}/\gamma T_c$) is estimated to be 0.58, which is substantially smaller than the expected BCS value of 1.43 [32]. The small jump in specific heat indicating that the quasiparticles participating in the superconductivity condensation experience strong elastic scattering because inelastic scattering usually enhances the jump ratio [33, 34]. Besides, the residual $\gamma_0$ value obtained by the extrapolation of $C_p/T$ curves linearly to 0 K as shown in Fig. 6 is $\gamma_0$ ~ 10.77 mJ/mol/K$^2$ which is about 71 % of $\gamma$, also supports the presence of strong scatterers [35]. The large $\gamma_0$ value indicates a large residual density of states at the Fermi energy $E_F$. The phenomenological theories suggest that only the Fermi surface of the majority band opens the superconducting gap and the minority band remains a normal state below $T_c$. And the larger value of $\gamma_0/\gamma$ in the case of Ta$_4$CoSi might indicate that the present superconductivity is extremely sensitive to a small amount of impurity. It can be seen from the PXRD refinement that there is a small part of *I4/mcm* phase in Ta$_4$CoSi alloy (see in **Fig. 1a**). And previous studies have demonstrated that the *I4/mcm* space group is a non-superconducting phase in the same CuAl$_2$-type compounds [19,20]. Furthermore, the non-superconducting contribution may also originate from defects in arc-melted material. Similar behaviors have been also observed in other systems [20,36-39], such as ternary silicide intermetallic compounds (Nb$_4$NiSi and URu$_2$Si$_2$) [20,36], transition metal chalcogenide superconductor (Nb$_2$Pd$_{0.81}$S$_5$) [37], and Fe-based superconductors (LiFeAs, NdFeAsO$_{1-x}$F$_x$) [38,39]. The Debye temperature ($\Theta_D$) of Ta$_4$CoSi was then calculated to be 346 K using the $\beta$ value in the formula $\Theta_D = (12\pi^4 nR/5\beta)^{1/3}$, where $R$ represents the ideal gas constant and $n$ represents the number of atoms per formula unit ($n = 6$). Moreover, assuming $\mu^* = 0.13$, the electron-phonon coupling constant ($\lambda_{ep}$) is further estimated to be 0.51, following the inverted McMillan equation $\lambda_{ep} = \frac{1.04 + \mu^* \ln\left(\frac{\Theta_D}{1.45 T_c}\right)}{(1 - 0.62\mu^*) \ln\left(\frac{\Theta_D}{1.45 T_c}\right) - 1.04}$ [40]. Finally, **Table 2** summarizes all the superconducting parameters.

As is shown in **Fig. 7**, the projected and total density of states (PDOS) of Ta$_4$CoSi was calculated by First-principles. The density of states (DOS) passing through the Fermi level demonstrates its metallic properties. PDOS diagram shows that the 5*d* electron of Ta is the main contribution near the Fermi level although the 3*d* electron of Co has some contribution. These results indicate that the superconductivity may mainly originate from the 5*d* electrons of Ta.

The new Ta$_4$CoSi superconductor has unique and elegant honeycomb Ta networks in the (110) plane, as displayed in **Fig. 1(b)**. And the First-principal calculations indicate that the superconductivity may mainly originate from the 5*d* electrons of Ta. Thus, we compare the new Ta$_4$CoSi superconductor with other hexagonal metal-chain-network superconductors and the remarkably high $T_c$ MgB$_2$ superconductor with the B-B honeycomb network. These compounds have a single atom honeycomb network structure. **Table 3** summarizes the parameters associated with superconductivity and the structures of hexagonal metal-chain-network-based compounds and $MgB_2$ superconductor. The $T_c$ as a function of the honeycomb lattice in honeycomb network superconductors is shown in **Fig. 8**. Interestingly, the $T_c$ is roughly linearly related to the honeycomb lattice. The new $Ta_4CoSi$ superconductor follows the relationship between $T_c$ and the honeycomb lattice function. $T_c$ gradually decreases from $MgB_2$ with B honeycomb network to $KV_3Sb_5$ with Sb honeycomb network as the honeycomb lattice from 1.781 to 3.160 Å. At the same time, the $\Theta_D$ decreases with the increase of the honeycomb lattice, which is displayed in the illustration in **Fig. 8**. The phenomenon that the $T_c$ is negatively correlated with the bond length is also observed in Nb-based superconductors [20]. Besides, the shrinkage of the Fe honeycomb lattice of layered $FePX_3$ ($X$ = S and Se) under high pressure leads to the emergency of SC and the increase of $T_c$ [42]. Thus, reducing the honeycomb lattice parameter may effectively increase the $T_c$ and $\Theta_D$ in the honeycomb network superconducting materials. It will be worth searching for more new superconductors with a honeycomb network and confirming this relationship.

**IV. Conclusion**

In conclusion, we have studied the crystal structure and its superconducting properties of the novel superconductor, $Ta_4CoSi$, prepared by the arc melting method. The $Ta_4CoSi$ compound adapts a $CuAl_2$-type structure (space group *P4/mcm*, No. 124), featuring the honeycomb networks of Ta-Ta in the (110) plane. Our studies show that $Ta_4CoSi$ is a type-II superconductor with $T_c$ ~ 2.45 K, $\mu_0H_{c1}(0)$ ~ 9.86 mT, and $\mu_0H_{c2}(0)$ ~ 0.84 T. This work shows that reducing the honeycomb lattice parameter (adding a pressure) may be an effective method to increase the $T_c$ in the honeycomb network superconducting materials. Furthermore, $Ta_4CoSi$ is a new platform for studying unique crystal structure and superconductivity, which could lead to insights into the physics of

Kagome superconductors, high-$T_c$ superconductors, and topological superconductors [43,44].

## Acknowledgments

This work is supported by the NSFC-11922415, NSFC-12274471, NKRDPC-2017YFA0206203, NKRDPC-2018YFA0306001, NSFC-11974432, NSFC-92165204, NSFC-U2032213, Guangdong Basic and Applied Basic Research Foundation (2022A1515011168, 2019A1515011718, 2019A1515011337), the Fundamental Research Funds for the Central Universities (19lgzd03), the Key Research & Development Program of Guangdong Province, China (2019B110209003), Leading Talent Program of Guangdong Special Projects (201626003), and the Pearl River Scholarship Program of Guangdong Province Universities and Colleges (20191001). J. S thanks the project funded by the IOP Startup Funds (BR202118). G.T.L thanks the project funded by China Postdoctoral Science Foundation (Grant No. 2022T150414).

**Figure captions:**

FIG. 1. Crystal structure characterization of Ta$_4$CoSi. (a) Rietveld refinement profile of the powder XRD of new superconducting phase Ta$_4$CoSi. (b) The crystal structure of Ta$_4$CoSi with space group *P4/mcc*. The Ta atoms form a honeycomb network in the (110) plane.

FIG. 2 The SEM (a), BSE (b), EDXS spectrum (c), and mapping of Ta$_4$CoSi sample.

FIG. 3 Temperature dependence of normalized $\rho/\rho_{300K}$ of Ta$_4$CoSi sample. The inset shows the low-temperature behavior in the vicinity of T$_c$.

FIG. 4 (a) Temperature-dependent ZFC magnetization under the field of H = 20 Oe, exhibiting a transition to diamagnetic Meissner state below 2.44 K. The volume susceptibility $\chi_v$ is corrected with the demagnetization factor derived from M versus H measurements. (b) Field-dependent magnetization curves measured at 1.80 - 2.25 K. (c) The M-M$_{fit}$ curves as a function of the magnetic fields. (d) Lower critical field estimation.

FIG. 5 (a) The magnetic field dependence of the superconducting transition for 0 T $\leq \mu_0$H $\leq$ 0.26 T. The 50 %-criterion is depicted as the black dashed line. (b) The $\mu_0$H-T phase diagram.

FIG. 6 Normalized electronic specific heats plotted a function of T/T$_c$. The inset displays the C$_p$/T versus T$^2$ for Ta$_4$CoSi at 0.6 to 10 K.

FIG. 7 Total and projected density of states (PDOS) calculated for Ta$_4$CoSi, black dotted lines indicate Fermi level.

FIG. 8 The honeycomb lattice parameter dependence of the *T$_c$*. The inset shows the honeycomb lattice dependence of the $\Theta_D$.

**Figures**

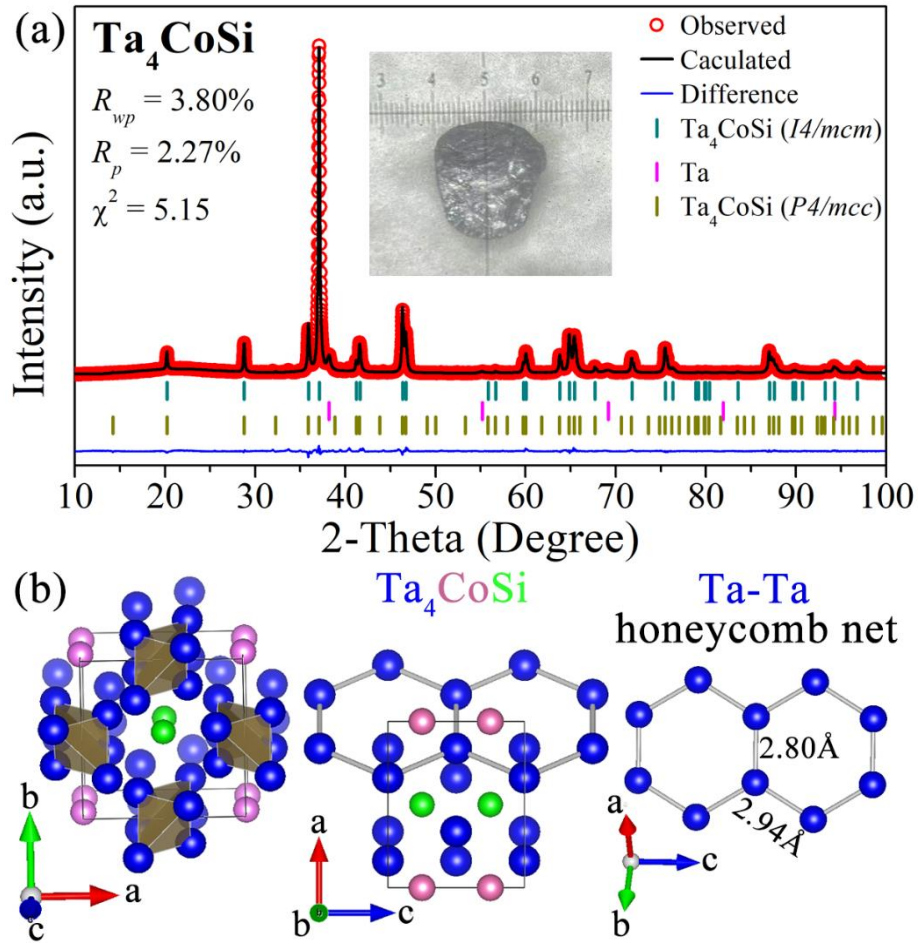

FIG. 1 Crystal structure characterization of Ta$_4$CoSi. (a) Rietveld refinement profile of the powder XRD of new superconducting phase Ta$_4$CoSi. (b) The crystal structure of Ta$_4$CoSi with space group *P4/mcc*. The Ta atoms form a honeycomb network in the (110) plane.

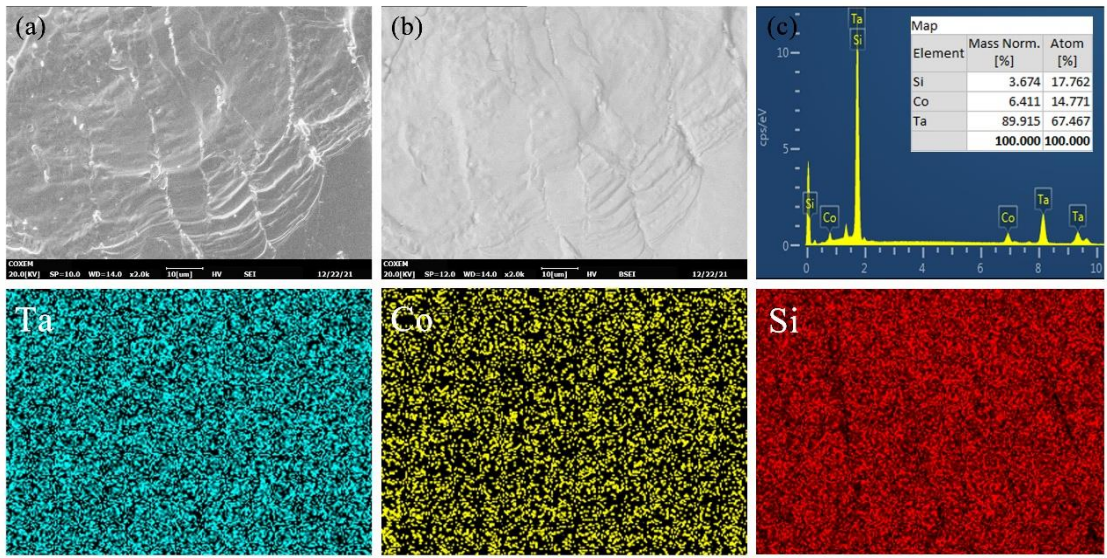

FIG. 2 The SEM (a), BSE (b), EDXS spectrum (c), and mapping of Ta$_4$CoSi sample.

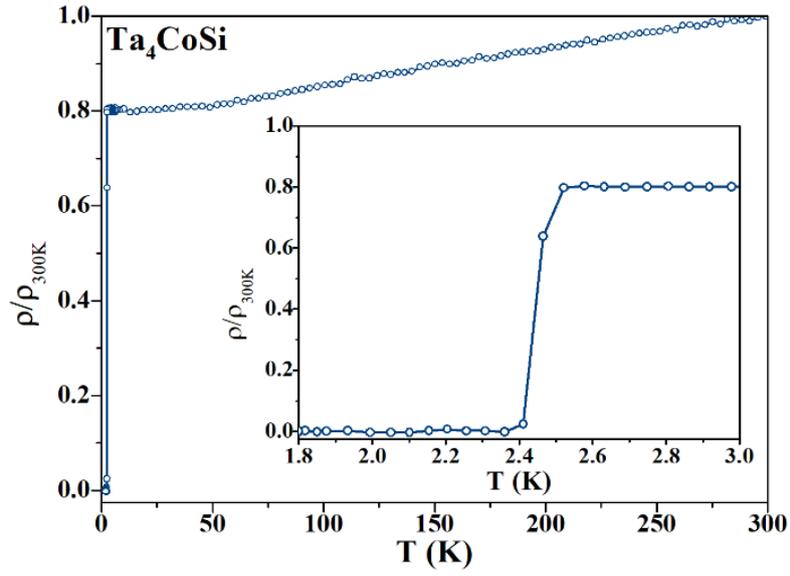

FIG. 3 Temperature dependence of normalized $\rho/\rho_{300K}$ of Ta$_4$CoSi sample. The inset shows the low-temperature behavior in the vicinity of $T_c$.

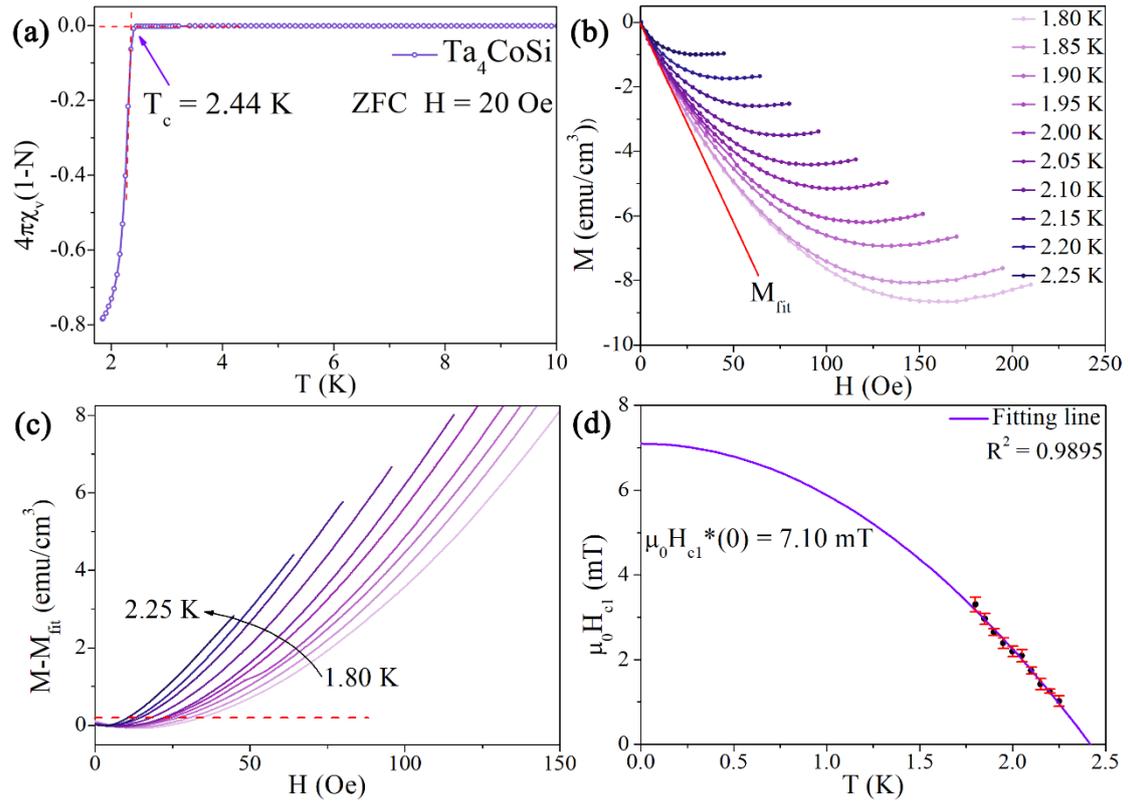

FIG. 4 (a) Temperature-dependent ZFC magnetization under the field of H = 20 Oe, exhibiting a transition to diamagnetic Meissner state below 2.44 K. The volume susceptibility $\chi_v$ is corrected with the demagnetization factor derived from M versus H measurements. (b) Field-dependent magnetization curves measured at 1.80 - 2.25 K. (c) The $M-M_{fit}$ curves as a function of the magnetic fields. (d) Lower critical field estimation.

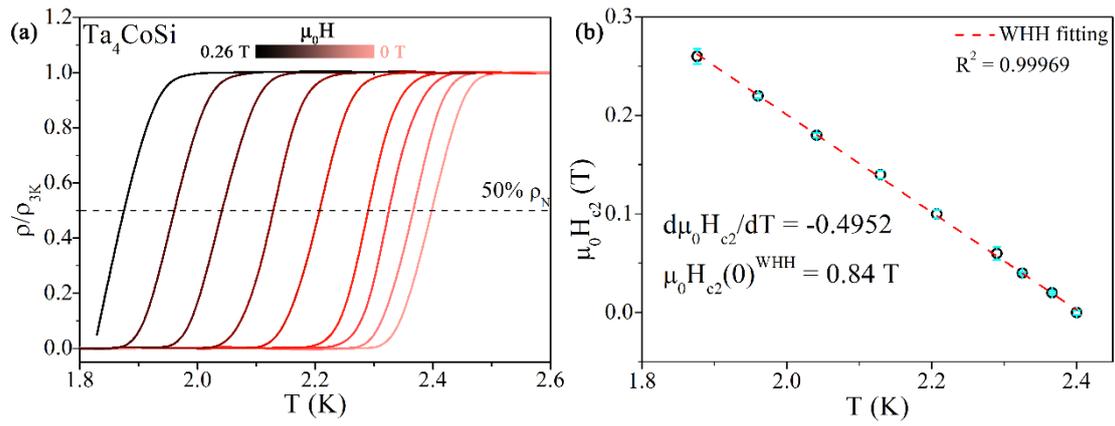

FIG. 5 (a) The magnetic field dependence of the superconducting transition for 0 T ≤ $\mu_0H$ ≤ 0.26 T. The 50 %-criterion is depicted as the black dashed line. (b) The $\mu_0H$-T phase diagram.

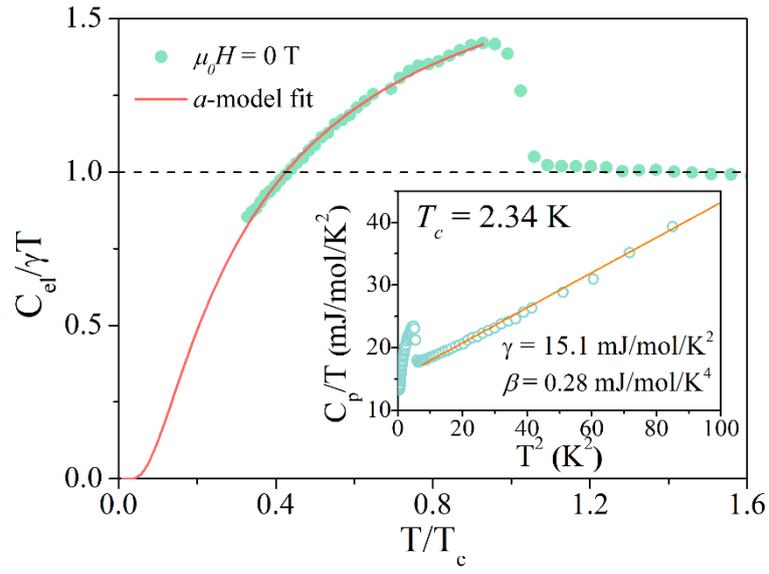

FIG. 6 Normalized electronic specific heats plotted a function of $T/T_c$. The inset displays the $C_p/T$ versus $T^2$ for $Ta_4CoSi$ at 0.6 to 10 K.

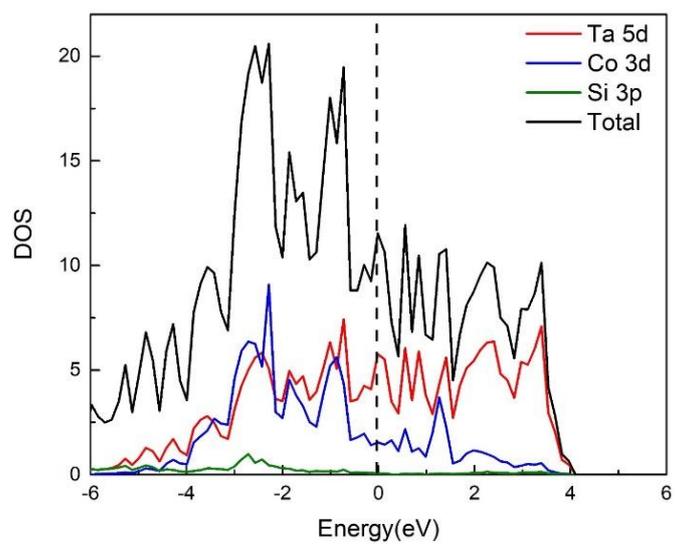

FIG. 7 Total and projected density of states (PDOS) calculated for $Ta_4CoSi$, black dotted lines indicate Fermi level.

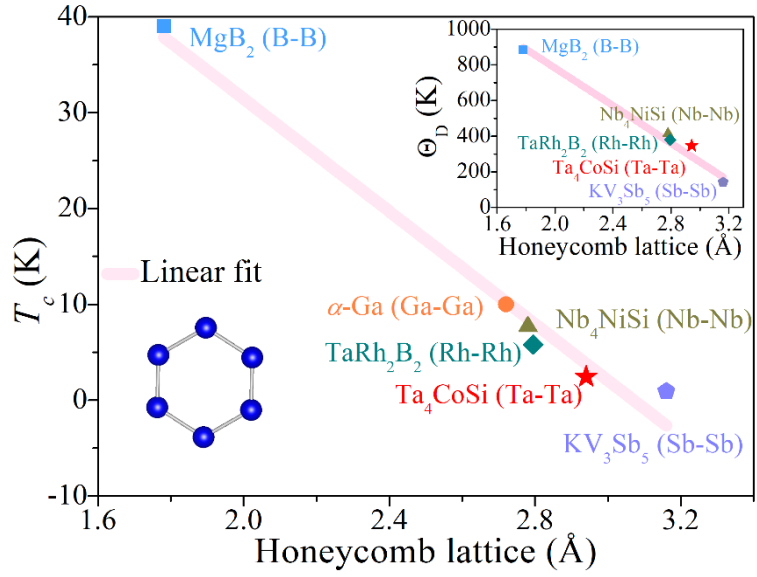

FIG. 8 The honeycomb lattice parameter dependence of the $T_c$. The inset shows the honeycomb lattice dependence of the $\Theta_D$.

Table 1. Structural parameters of superconducting Ta$_4$CoSi at 300 K.

| Atom | x | y | z | Wyckoff | Occupancy |
|---|---|---|---|---|---|
| Ta | 0.16233(2) | 0.65777(2) | 0 | 8$m$ | 1.0 |
| Co | 0 | 0 | 0.25 | 2$a$ | 1.0 |
| Si | 0.5 | 0.5 | 0.25 | 2$c$ | 1.0 |

The space group is *P4/mcc* (No.124), $a$ = b = 6.17755(7) Å, $c$ = 4.98443(8) Å.

Table 2. Observed superconducting parameters of Ta$_4$CoSi.

| parameter | unit | value |
|---|---|---|
| $T_c$ | K | 2.45 |
| $\mu_0 H_{c1}(0)$ | mT | 9.86 |
| $\mu_0 H_{c2}(0)$ | T | 0.84 |
| $\mu_0 H_c(0)$ | mT | 26.99 |
| $\mu_0 H^{Pauli}$ | T | 4.53 |
| $\Delta C/\gamma T_c$ |  | 0.58 |
| $\Theta_D$ | K | 346 |
| $\lambda_{ep}$ |  | 0.51 |

Table 3. Comparison between the superconducting and structural parameters among honeycomb metal-chain-network superconductors and MgB$_2$ superconductor (honeycomb network of B atoms).

|  | MgB$_2$ [28] | $a$-Ga [41] | Nb$_4$NiSi [20] | TaRh$_2$B$_2$ [29] | Ta$_4$CoSi [This work] | KV$_3$Sb$_5$ [30] |
|---|---|---|---|---|---|---|
| $T_c$ (K) | 39 | 10 | 7.7 | 5.8 | 2.45 | 0.93 |
| Honeycomb lattice (Å) | 1.781 | 2.720 | 2.779 | 2.795 | 2.943 | 3.160 |
| $\Theta_D$ (K) | 884 | - | 415 | 379 | 346 | 141 |
| $\lambda_{ep}$ | 0.680 | - | 0.600 | 0.624 | 0.510 | 0.460 |